\documentclass[11pt]{article}
\usepackage[margin=1in]{geometry}
\usepackage{graphicx,caption}
\graphicspath{{Fig/}}
\usepackage{microtype}
\usepackage{amsmath, amssymb}
\usepackage{amsthm}
\usepackage[hidelinks]{hyperref}           
\usepackage{authblk}
\usepackage[hidelinks]{hyperref}
 \date{}
\theoremstyle{plain}

\theoremstyle{definition}

\theoremstyle{remark}

\usepackage{authblk}
\title{SCUDDO: Single-cell Clustering Using Diagonal Diffusion Operators}
\author[1,*]{Luka Maisuradze}
\author[2]{Mark D. Shattuck}
\author[3,4]{Corey S. O'Hern}
\affil[1]{{Department of Molecular Biophysics and Biochemistry}, {Yale University}}
\affil[2]{{Benjamin Levich Institute and Physics Department}, {The City College of New York}}
\affil[3]{{Department of Mechanical Engineering}, {Yale University}}
\affil[4]{{Graduate Program in Computational Biology and Biomedical Informatics}, {Yale University}}
\affil[*]{\textit{Corresponding author:} \href{mailto:lgmaisura@gmail.com}{lgmaisura@gmail.com}}

\begin{document}
\maketitle

\abstract{

\textbf{Motivation:} Advances in high-throughput chromatin conformation capture have provided insight into the three-dimensional structure and organization of chromatin. While bulk Hi-C experiments capture spatio-temporally averaged chromatin interactions across millions of cells, single-cell Hi-C experiments report on the chromatin interactions of individual cells. Supervised and unsupervised algorithms have been developed to embed single-cell Hi-C maps and identify different cell types. However, single-cell Hi-C maps are often difficult to cluster due to their high sparsity, with state-of-the-art algorithms achieving a maximum Adjusted Rand Index (ARI) of only $\lesssim 0.4$ on several datasets while requiring labels for training.  \\

\textbf{Results:} We introduce a novel unsupervised algorithm, Single-cell Clustering Using Diagonal Diffusion Operators (SCUDDO), to embed and cluster single-cell Hi-C maps. We evaluate SCUDDO on three previously difficult-to-cluster single-cell Hi-C datasets, and show that it can outperform other current algorithms in ARI by $\gtrsim 0.2$. Further, SCUDDO outperforms all other tested algorithms even when we restrict the number of intrachromosomal maps for each cell type and when we use only a small fraction of contacts in each Hi-C map. Thus, SCUDDO can capture the underlying latent features of single-cell Hi-C maps and provide accurate labeling of cell types even when cell types are not known {\it a priori}. \\

\textbf{Availability:} SCUDDO is freely available at www.github.com/lmaisuradze/scuddo. The tested datasets are publicly available and can be downloaded from the Gene Expression Omnibus.}

\maketitle

\section{Introduction}\label{sec1}
Elucidating the structure and dynamics of chromatin in cell nuclei is essential for understanding numerous cellular processes such as DNA transcription and replication ~\cite{bib87}. Advances in whole-genome analyses, e.g. chromosome conformation capture techniques such as Hi-C, have provided important insights into long-ranged chromatin interactions and hierarchical chromatin organization ~\cite{bib86}. Hi-C experiments provide chromatin contact maps, often represented as a symmetric matrix $\mathcal{A}$, where $\mathcal{A}_{ij}$ gives the number of times that loci $i$ and $j$ of chromatin come into close proximity. Bulk Hi-C contact maps provide information on chromatin fragment interactions averaged over millions of cells. In contrast, $\textit{single-cell}$ Hi-C maps give the frequency of chromatin contacts in each individual cell.

It is now well established that chromatin structure and organization can differ significantly across cell populations~\cite{bib87,bib88}. Transcription analyses and imaging studies have shown that gene expression profiles and cell morphology can differ even between genetically identical cells~\cite{bib88,bib89,bib90}. In addition, the size and location of chromatin loops and topologically associating domains (TADs) can vary between the Hi-C maps of individual cells for a given organism ~\cite{bib91,bib92,bib93,bib94}. As a result, the loci that posses high contact frequencies in bulk Hi-C maps can differ from those that are in close spatial proximity in fluorescence in situ hybridization (FISH) experiments, in part due to the heterogeneity in chromatin structure across individual cells ~\cite{bib95,bib96,bib97,bib98,bib99}. Thus, bulk Hi-C maps cannot be used to capture the structure and organization of chromatin in individual cells.
 
\begin{figure}[!t]%
\centering
\captionsetup{width=.9\linewidth}
\includegraphics[scale=.75]{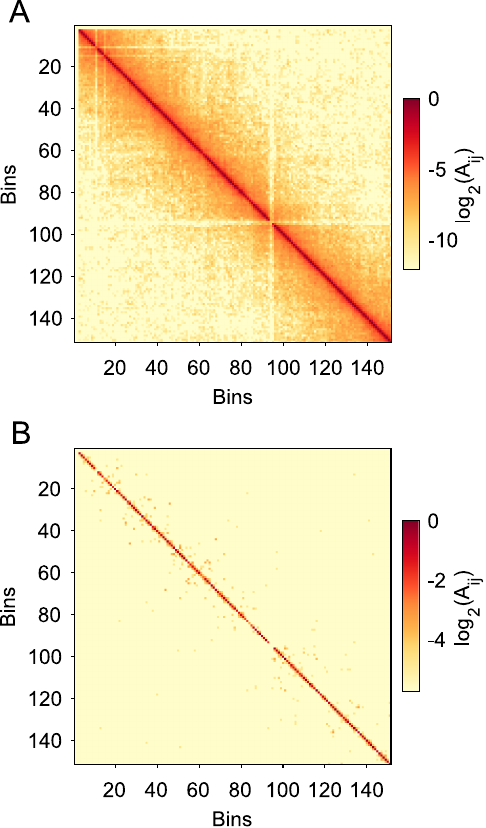}
\caption{(A) A psuedo-bulk Hi-C map ($\log_2 \mathcal{A}_{ij}$) for chromosome $5$ using pooled mouse oocyte cells before division from the Collombet \cite{bib118}  dataset (normalized so that $\text{max}(\mathcal{A}_{ij}) =1$). (B) An example single-cell Hi-C map from the Collombet dataset for chromosome $5$ of a mouse oocyte cell using the same normalization.}
\label{fig1}
\end{figure}

Several single-cell Hi-C technologies have been developed to capture chromatin interactions for large numbers of individual cells ~\cite{bib100,bib101,bib102, bib103,bib104}. Single-cell Hi-C techniques enable studies of genome organization in individual cells, as well as comparisons of chromatin structure and organization across different cell types. Using data from single-cell Hi-C experiments, computational studies have focused on TAD, loop, and compartment identification for individual cells within a population ~\cite{bib103,bib104}. However, despite the rapid advances in genome-wide assays, single-cell Hi-C maps are still sparse, only capturing a fraction of the interactions that are obtained in bulk Hi-C experiments ~\cite{bib105,bib106}. For example, in Fig.~\ref{fig1}, we show a psuedo-bulk Hi-C map for chromosome $5$ in mouse oocyte cells~\cite{bib118} and compare it to a single-cell Hi-C map for the same chromosome and cell type. The single-cell Hi-C map shows significant sparsity, with most of the off-diagonal elements having a count of $0$, as well as large variability for elements near the diagonal. 

While techniques like fluorescence-activated cell sorting can be used to label single cells during chromosome conformation capture methods, these techniques are more expensive, lower throughput, and not as widely available as single-cell Hi-C experiments. Thus, the development of classification algorithms for single-cell Hi-C maps may enable researchers to identify the key chromatin interactions that distinguish different cell types. Algorithms developed for bulk Hi-C analysis, including topologically associating domain callers ~\cite{bib130}, often work with limited efficacy on raw single-cell Hi-C data. Thus, due to their inherent variability and sparsity, specialized algorithms must be developed to identify robust features in single-cell Hi-C maps. In this article, we focus on the specific task of classifying single-cell Hi-C maps based on the cell labels that have been provided by the experimental studies. Most algorithms for clustering single-cell Hi-C maps use dimensionality reduction, treating each single-cell Hi-C map as a point in high-dimensional space and then mapping each point to a lower-dimensional space to cluster the data~\cite{bib105,bib106,bib107}. Despite the fact that there are more than a dozen algorithms to date for clustering single-cell Hi-C maps, there are many single-cell Hi-C datasets for which these methods achieve a maximum adjusted Rand index ${\rm ARI} \lesssim 0.4$~\cite{bib106,bib133}. Moreover, there are many cases where one clustering method performs well on one single-cell Hi-C dataset, but then performs poorly on another dataset~\cite{bib105,bib106}, suggesting that current methods have trouble identifying features that generalize across multiple single-cell Hi-C datasets for clustering.

We develop a novel algorithm, SCUDDO (single-cell clustering using diagonal diffusion operators), which is fully unsupervised, fast, and easy to interpret to separate single-cell Hi-C maps into distinct groups. We then compare the predicted labels of the single-cell Hi-C maps to the cell types that are provided by experimental studies. To quantify the accuracy of the clustering, we calculate the ARI and normalized mutual information (NMI) using the predicted and ground truth labels. We find that SCUDDO outperforms current state of the art methods on three difficult-to-cluster single-cell Hi-C datasets, achieving an ARI and NMI greater than those for all of the tested methods on each of the datasets. We also find that SCUDDO achieves higher accuracy for clustering single-cell Hi-C maps compared to other algorithms when using only a fraction of the number of intrachromosomal maps and a fraction of the diagonals in each map.

The remainder of the manuscript is organized as follows. In the Materials and Methods section, we describe the key elements and hyperparameters of SCUDDO for clustering single-cell Hi-C maps. We then describe the three difficult-to-cluster datasets for benchmarking the new algorithm and the two metrics (ARI and NMI) for quantifying the clustering accuracy. In the Results section, we provide the ARI and NMI scores for SCUDDO and three current methods for clustering single-cell Hi-C maps on each of the three difficult-to-cluster Hi-C datasets. We also show SCUDDO's performance across different hyperparameter regimes and when limiting the number of diagonals and intrachromosomal maps sampled. In the Discussion, we include some interpretations of the results,  our conclusions, and promising future research directions.

\begin{figure*}[!t]%
\centering
\captionsetup{width=.9\linewidth}
\includegraphics[scale=.75]{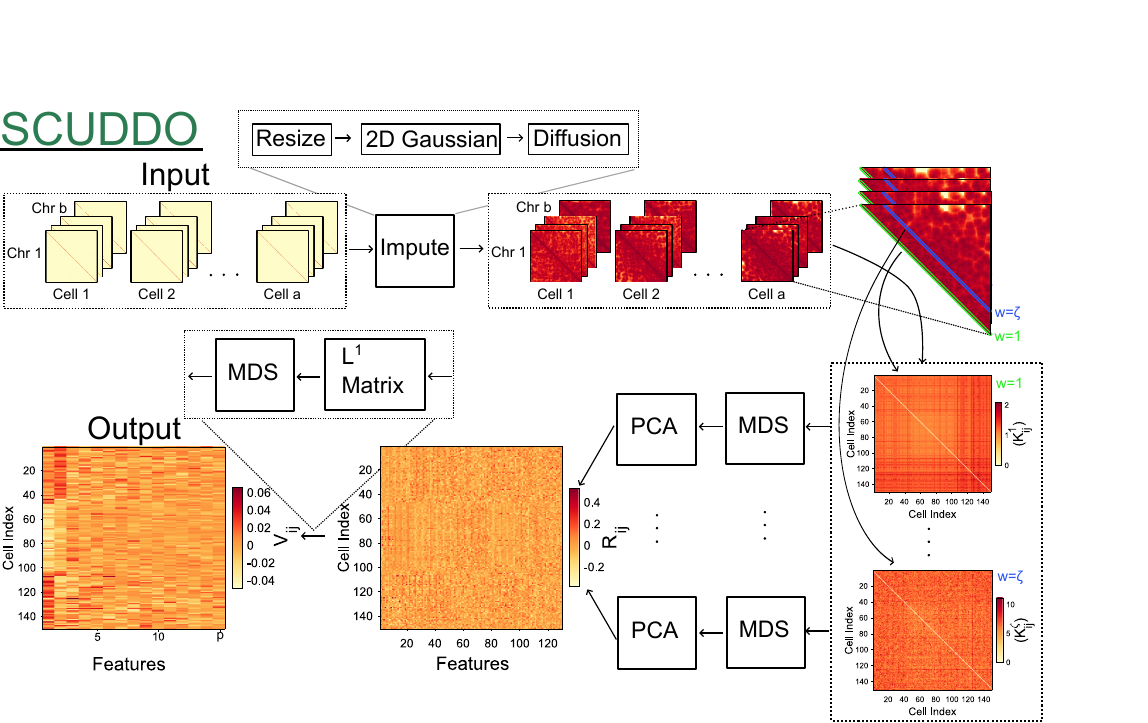}
\caption{A schematic of the SCUDDO algorithm for clustering single-cell Hi-C maps. We illustrate the method using intrachromosomal Hi-C maps from the Li, et al. dataset~\cite{bib118}. SCUDDO first imputes the set of intrachromosomal Hi-C matrices (indexed by $\mathcal{A}^k_{s}$) for each cell and then samples each diagonal (indexed by $w$) from each Hi-C matrix to form a feature matrix $\mathcal{K}^w$ for each sampled diagonal. Principal component analysis (PCA) and nonmetric multi-dimensional scaling (MDS) are then applied to each feature matrix to form the matrix $\mathcal{R}$, which is then embedded in a lower dimensional latent space using the $L^1$ norm to form the embedding, $\mathcal{V}$.}
\label{fig2}
\end{figure*}

\section{Materials and Methods}\label{sec2}

The Materials and Methods is organized into three subsections. We first define the necessary notation and summarize the steps of the SCUDDO method to cluster single-cell Hi-C maps. Second, we describe three difficult-to-cluster single-cell Hi-C datasets that will be used to benchmark SCUDDO alongside three other current algorithms. Finally, we define the two metrics, ARI and NMI, which are used to quantify the unsupervised clustering accuracy.

\subsection{SCUDDO algorithm}\label{subsec1}

The SCUDDO algorithm takes as input a set of intrachromosomal Hi-C maps with nonnegative integer entries for $a$ cells, each with $b$ chromosomes, totaling $a\times b$ intrachromosomal Hi-C maps. As for bulk Hi-C maps, single-cell Hi-C maps are represented as symmetric matrices with elements $\mathcal{A}_{ij}$ that represent the number of contacts between loci $i$ and $j$ on chromatin. To distinguish between the cell and chromosome indices, we define $\mathcal{A}^k_{s,ij}$ as the $ij\text{th}$ element of the $n^k \times n^k$ Hi-C map for chromosome $k$ of cell $s$. $n^k$ only depends on $k$ since the dimensions of the Hi-C map only vary across different chromosomes. Given a set of intrachromosomal matrices, SCUDDO returns a low-dimensional embedding of the Hi-C matrices. This embedding is then used as an input into a clustering algorithm, for example K-means++ ~\cite{bib120}, where each cell is assigned to one of $l$ predicted labels.

SCUDDO starts by pre-processing and performing imputation on each intrachromosomal matrix $\mathcal{A}^k_s$. First, each $\mathcal{A}^k_s$ is reshaped into the same size $r \times r$ matrix, $\mathcal{A}'^k_s$, using a bicubic interpolation kernel, where $r=\sum_{k=1}^{b} n^k/b$. SCUDDO then convolves each intrachromosomal matrix with a Gaussian kernel:

\begin{equation}
\mathcal{A}''^k_{s,ij}=  \sum_{\chi=1}^{9} \sum_{\omega=1}^{9} 
\mathcal{G}_{\chi \omega} \mathcal{A}'^{k}_{s,(i-{4} + \chi) (j-4 + \omega)},
\end{equation} 
where $\mathcal{G}$ is a two-dimensional $9 \times 9$ Gaussian kernel with standard deviation $\sigma=0.5$ that uses replicate padding, where values outside of the bounds of the original Hi-C map are set to the values of the nearest border entry. $\mathcal{G}$ smooths local regions in each individual intrachromosomal matrix. The final pre-processing step is to normalize each intrachromosomal matrix and apply a diffusion kernel via a matrix exponential:

 \begin{equation}
\mathcal{\mathcal{B}}^k_{s}= \text{exp}\left(-\frac{\mathcal{A}''^k_{s}}{\sum_{ij} \mathcal{A}''^k_{s,ij}}  \right),
\end{equation} 
which represents backwards diffusion over $\mathcal{A}''^k_{s}$.
Next, we construct high-dimensional embeddings of the intrachromosomal Hi-C maps for each cell. Let $d^w(\mathcal{B}^k_s)$ be the ordered set of Hi-C map entries on the $w$th superdiagonal of the $r \times r$ matrix $\mathcal{B}^k_s$:
\begin{equation}
d^w(\mathcal{B}^k_s) = \{\mathcal{B}^k_{s,1,1+w}, \mathcal{B}^k_{s,2,2+w},\ldots, \mathcal{B}^k_{s,r-w,r} \}.
\end{equation}
For a given $w$, $d^w(\mathcal{B}^k_s)$ for each chromosome $k$ for cell $s$ is concatenated to form the embedding vector:
\begin{equation}
  {\vec e}^{~w}_s  = \{d^w(\mathcal{B}^1_s) ,  d^w(\mathcal{B}^2_s) ,\ldots, d^w(\mathcal{B}^b_s) \},
\end{equation} 
where $\alpha=1,\ldots,b(r-w)$ indexes the entries in ${\vec e}^{~w}_s$. Every embedding vector, ${\vec e}^{~w}_s $, is z-score normalized such that 
\begin{equation}
  {\vec e}_s~'^{w}  = \frac{{\vec e}^{~w}_s - \mu}{\sqrt{\frac{1}{b(r-w)-1}\sum_{\alpha=1}^{b(r-w)} |{\vec e}^{~w}_s - \mu|}},
\end{equation} 
where $\mu= \sum_{\alpha=1}^{b(r-w)}(\vec e^{~w}_s)_\alpha/[b(r-w)]$. This pooling approach is similar to previous work~\cite{bib133,bib134} that employs band normalization for Hi-C matrices. Next, each embedding vector is transformed into a signed difference vector:
\begin{equation}
\label{ternary}
 {\vec f}^{~w}_s  = \text{sgn}(\nabla ({\vec e}~'^{~w}_s)),
\end{equation} 
where $\nabla({\vec e}~'^{~w}_s)_{\alpha}=({\vec e}~'^{~w}_s)_{\alpha}-({\vec e}~'^{~w}_s)_{\alpha+1}$ and sgn is the sign function. (Note that we set the last entry of the chromosome difference vector $({\vec f}^{~w}_s)_{b(r-w)}= ({\vec e}~'^{~w}_s )_{b(r-w)}$.) Eq.~\ref{ternary} transforms ${\vec e}~'^{~w}_s$ into a ternary vector with values $1$, $0$, or $-1$. For a given $w$, each cell's difference vector,  ${\vec f}^{~w}_1,{\vec f}^{~w}_2,\ldots, {\vec f}^{~w}_a$ is used to calculate the distance matrix between cells $i$ and $j$ using cosine similarity:
\begin{equation}
\label{distance}
 D^w_{ij}  = 1-  \frac{ {\vec f}^{~w}_i  \cdot  {\vec f}^{~w}_j }{|{\vec f}^{~w}_i| |{\vec f}^{~w}_j| },
\end{equation}
where $|{\vec X}|$ indicates the magnitude of ${\vec X}$. 
A separate distance matrix is calculated for ${\vec e}~'^{~w}_s$:
\begin{equation}
\label{distance}
 D'^w_{ij}  = 1-  \frac{ {\vec e}~'^{~w}_i  \cdot  {\vec e}~'^{~w}_j }{|{\vec e}~'^{~w}_i| |{\vec e}~'^{~w}_j| },
\end{equation}
and combined to form a final exponentiated distance matrix using element-wise exponentiation:
\begin{equation}
\label{distance}
 \mathcal{K}^w_{ij}  = e^{(D'^w_{ij} +D^w_{ij}) (D'^w_{ij} D^w_{ij})} .
\end{equation} 
Finally, SCUDDO uses nonmetric multidimensional scaling (MDS)~\cite{bib129} to transform the $a\times a$ matrix $K^w$ into a lower dimensional representation, i.e. an $a\times p$ matrix where $p <a$, which preserves the distances in $\mathcal{K}^w$. The multidimensional scaling is followed by principal component analysis to further reduce the dimension to an $a \times q$ matrix $\mathcal{U}^w$, where $q<p$ ($p=30$ and $q=5$).  This procedure is performed for the diagonal ($w=0$) and a given number of superdiagonals ($w=\zeta>0$), and each set of dimensionality-reduced representations are concatenated, forming the $a \times (q(\zeta+1))$ matrix $\mathcal{R}={\mathcal{U}^0, \mathcal{U}^1,\ldots,\mathcal{U}^\zeta}$.  $\mathcal{R}$ is then normalized feature-wise using the softmax function, $\mathcal{R}'_{ij}=e^{\mathcal{R}_{ij}}/\sum_{\theta=1}^a e^{\mathcal{R}_{\theta j}}$, and a distance matrix is constructed using the $L^1$ metric: 
\begin{equation}
\label{distance}
 \mathcal{S}_{ij}  = \sum^{qw}_{\lambda=1} |\mathcal{R'}_{i\lambda}-\mathcal{R'}_{j\lambda}| .
\end{equation} 
Another round of dimensionality reduction is performed using multidimensional scaling to reduce the dimension of $\mathcal{S}$ to the embedding size $\epsilon$, which gives the $a \times \epsilon$ matrix, $\mathcal{V}$. Because there is no guarantee of convexity associated with each cluster when clustering single cell Hi-C matrices, we use spectral decomposition before performing the clustering. In particular, SCUDDO transforms $\mathcal{V}$ into the similarity matrix, $\mathfrak{A}_{ij}=e^{-\mathcal{Z}_{ij}^2}$, where $\mathcal{Z}_{ij} =  |\vec{\mathcal{V}}_{i*} - \vec{\mathcal{V}}_{j*}|$ and $\mathcal{V}_{i*}$ is the vector consisting of all elements in the $i^{\text{th}}$ row of $\mathcal{V}$. Next, we calculate the final $a \times l$ spectral embedding $\mathcal{C}$, where the columns of $\mathcal{C}$ are the smallest $l$ eigenvectors of the random-walk Laplacian matrix constructed from $\mathfrak{A}$ using the Shi-Malik algorithm \cite{bib135} with $\log(a)$ nearest neighbors. We then input $\mathcal{C}$ and the number of labels $l$ into a clustering algorithm, such as K-means++ ~\cite{bib120}, which returns the predicted labels for each single-cell Hi-C map. 

SCUDDO includes two tunable hyperparameters: $\zeta$, the set of (super)diagonals, $w=0, 1,\ldots,\zeta$ used to construct the embedding vectors, and the dimension $\epsilon$, to which $\mathcal{V}$ is reduced. By default, SCUDDO outputs two embeddings: $\mathcal{C}$ and $\mathcal{V}$. Both $\zeta$ and $\epsilon$ are varied in the Results section to study their effects on SCUDDO's performance for each dataset. For all  results in this study unless otherwise noted, we use $\mathcal{C}$ as the input into K-means++ and set $\zeta= 25$ and $\epsilon=5$. Our results are not sensitive to the values of the dimensions $p$ and $q$. 

 \subsection{Benchmarking of single-cell Hi-C clustering algorithms}\label{subsec2}

We focus our studies on three difficult-to-cluster datasets of single-cell Hi-C maps from recent benchmarking studies~\cite{bib106,bib128}. In particular, we consider the Li, et al.~\cite{bib101} dataset (GEO ID: GSE119171) consisting of $a=150$ mouse embryonic stem cells that are separated into $l=3$ labels: ``2i'', ``Serum1'', and ``Serum2'', the Flyamer, et al.~\cite{bib103} dataset (GEO ID: GSE80006) consisting of $a=134$ cells from developing mouse zygotes and oocytes with $l=3$ cell types: ``Oocyte'', ``ZygP'', and ``ZygM'' as labels, and the Collombet, et al.~\cite{bib118} dataset (GEO ID: GSE129029) consisting of $a=648$ mouse embryo cells with labels that represent $l=5$ different cell stages: $1$-cell, $2$-cell, $4$-cell, $8$-cell, and $64$-cell stages. In previous benchmarking studies \cite{bib106}, none of the eight tested methods for single-cell Hi-C map clustering achieved ARI or NMI $\geq 0.6$ on the Collombet, et al. dataset and in another study \cite{bib133}  none of the eight methods tested achieved an ARI $>0.45$ on the Li, et al. dataset across any clustering algorithm (not just k-means) . For each dataset, we use $1$ Mb bin sizes for the single-cell Hi-C maps, and re-bin those with higher resolution, as discussed in Zhou, et al.~\cite{bib125}. If the sum of all non-diagonal nonzero pairs of elements in the intrachromosomal Hi-C maps for a given cell is less than $5000$, the data for this cell was not included in the analysis. Also, for each individual chromosome of size $x$ for a cell, if the intrachromosomal Hi-C map for that chromosome has a sum of non-diagonal contacts that is less than $x$, all intrachromosomal Hi-C maps are not considered for that cell. 

After considering previous single-cell Hi-C map clustering studies~\cite{bib106, bib124, bib128, bib133}, we selected consistent top performers across several datasets to compare with SCUDDO: i.e. the Higashi~\cite{bib124}, HiCRep/MDS~\cite{bib132, bib107}, and scHiCluster~\cite{bib125} algorithms. While HiCRep/MDS is not as accurate as Higashi and scHiCluster, we include it in our analysis since it is the most widely used and best performing method that uses MDS similar to SCUDDO to the best of our knowledge. Importantly, all algorithms that we tested are unsupervised or self-supervised, and do not require labels for training. Other algorithms that require labels or significant pretraining are unable to cluster unlabeled single-cell Hi-C datasets, and thus they are not included in this manuscript. For each algorithm, we used the default hyperparameters and used the final embeddings (with no further processing) as input into K-means++ clustering to benchmark our calculations. 

\subsection{Metrics for clustering accuracy}\label{subsec3}

To assess the accuracy of the predicted labels, we calculate the adjusted rand index (ARI)~\cite{bib121} and normalized mutual information (NMI)~\cite{bib122}. Let  $\Omega_T(s)$ and  $\Omega_G(s)$ be functions that map each cell index $s$ (from $1$ to $a$) to the integers $l'$ and $l''$ respectively, where $l'$ is the ground truth label for cell $s$ and $l''$ is the predicted label for cell $s$.  We then define $P_T=\bigl\{X_1, X_2,...X_l\bigl\}$ as the ``ground-truth" label set, where $X_{l'}$ denotes the set of cells such that $\Omega_T(s)=l'$, and $P_G =\bigl\{Y_1, Y_2,...Y_l\bigl\}$ as the ``predicted'' label set, where $Y_{l''}$ is the set of cells such $\Omega_G(s)=l''$.

The adjusted Rand index determines the similarity between the sets of cells with given ground truth and predicted labels: 
\begin{equation}
\text{ARI} =  \frac{ \sum_{i=1}^{l}\sum_{j=1}^{l}{\beta_{ij} \choose 2}-  ({\sum_{i=1}^{l}{\Gamma_i \choose 2}   \sum_{j=1}^{l} {\Delta_{j} \choose 2}})/{a \choose 2}}        { \frac{1}{2} (\sum_{i=1}^{l}{\Gamma_i \choose 2} +  \sum_{j=1}^{l} {\Delta_j  \choose 2}) -  ({\sum_{i=1}^{l}{\Gamma_i \choose 2}   \sum_{j=1}^{l} {\Delta_j \choose 2}})/{a \choose 2}},
\end{equation} 
where $\beta_{ij}=[X_i \cap Y_j]$, $\cap$ is the intersection between two sets, $[X]$ is the number of elements in set $X$, $\Gamma_k = \sum_{i=1}^l \beta_{ki}$,  $\Delta_k = \sum_{j=1}^l \beta_{jk}$, and ${m\choose n}=\frac{m!}{n!(m-n)!}$.  ${\rm ARI}=1$ indicates a perfect match between $P_T$ and $P_G$, whereas ${\rm ARI}=0$ indicates the match between $P_T$ and $P_G$ is no better than that achieved by random assignments in $P_G$.

We also quantify the accuracy of the clustering of the single-cell Hi-C maps using the normalized mutual information (NMI). NMI measures how much information can be learned about a given clustering by observing a different, but related clustering. NMI is defined as:
\begin{equation}
\text{NMI} = \frac{ \sum_{i=1}^{l} \sum_{j=1}^{l} \mathcal{H}(i,j) \log_2 \frac{\mathcal{H}(i,j)}{\mathcal{H}(i)\mathcal{H}(j))}}
{\sqrt{(-\sum_{i=1}^l \mathcal{H}(i) \log_2 \mathcal{H}(i))(-\sum_{j=1}^l \mathcal{H}(j) \log_2 \mathcal{H}(j))}},
\end{equation}
where $\mathcal{H}(i)=\frac{[X_i]}{a}$, $\mathcal{H}(j)=\frac{[Y_j]}{a}$, and $\mathcal{H}(i,j)=\frac{[Y_i \cap X_j]}{a}$. $0 < {\rm NMI} < 1$, where ${\rm NMI}=1$ indicates that $P_T = P_G$ and ${\rm NMI}=0$ indicates that there is no correlation between $P_T$ and $P_G$.  We calculate both ARI and NMI since they can differ for different sized clusters: ARI is preferable when the sets in $P_T$ are similar in size, whereas NMI is preferable when the sets in $P_T$ are unbalanced.
For all datasets and algorithms, we calculate the ARI and NMI after using the native embedding and K-means++ clustering. 

\begin{figure}[!t]%
\centering
\includegraphics[scale=.90]{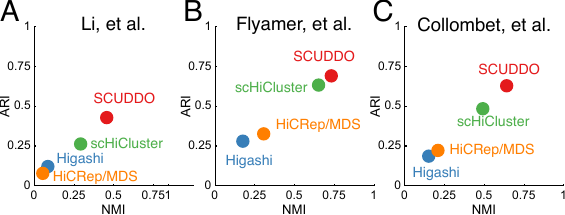}
\captionsetup{width=.9\linewidth}
\caption{Accuracy of the four single-cell Hi-C map clustering algorithms
(Higashi ~\cite{bib124} (blue), HiCRep/MDS ~\cite{bib132, bib107} (orange), scHiCluster ~\cite{bib125} (green), and SCUDDO (red)) on three difficult-to-cluster single-cell Hi-C datasets. We plot the adjusted Rand index (ARI) versus the normalized mutual information (NMI) for each algorithm on the (A) Li, et al. ~\cite{bib101}, (B) Flyamer, et al. ~\cite{bib103}, and (C) Collombet, et al.~\cite{bib118} datasets.}
\label{fig3}
\end{figure}

\section{Results}\label{sec3}

We carried out single-cell Hi-C map clustering on three difficult-to-cluster datasets (Collombet, et al.~\cite{bib118}, Flyamer, et al.~\cite{bib103}, and Li, et al.~\cite{bib101}) using three current algorithms (Higashi~\cite{bib124}, HiCRep/MDS~\cite{bib106}, and scHiCluster~\cite{bib125}) and compared the results to those obtained from SCUDDO. We plot ARI versus NMI for each dataset and algorithm in Fig.~\ref{fig3} (A)-(C). Overall, SCUDDO outperforms the other three methods for all datasets tested. For the Li, et al. dataset in Fig.~\ref{fig3} (A), we find a significant separation in accuracy between SCUDDO and the next most accurate method (scHiCluster).  SCUDDO achieves an ${\rm ARI} \sim 0.45$ and ${\rm NMI} \approx 0.42$, while scHiCluster achieves ${\rm ARI} \approx 0.29$ and ${\rm NMI} \approx 0.25$. For the Flyamer, et al. dataset in Fig.~\ref{fig3} (B), we find that SCUDDO has ${\rm ARI} \sim {\rm NMI} \approx 0.73$, whereas the next most accurate method, again SciHiCluster,
has ${\rm ARI} \sim {\rm NMI} \approx 0.65$. We also find that on some runs of K-means++ for this dataset, SCUDDO can achieve ${\rm ARI} \geq 0.90$. Lastly, for the Collombet, et al. dataset in Fig.~\ref{fig3} (C), we find that SCUDDO has ${\rm ARI} \sim {\rm NMI} \approx 0.64$, whereas the next most accurate method, again SciHiCluster, has ${\rm ARI} \sim {\rm NMI} < 0.5$.

\begin{figure*}[!t]%
\centering
\captionsetup{width=.9\linewidth}
\includegraphics[scale=1]{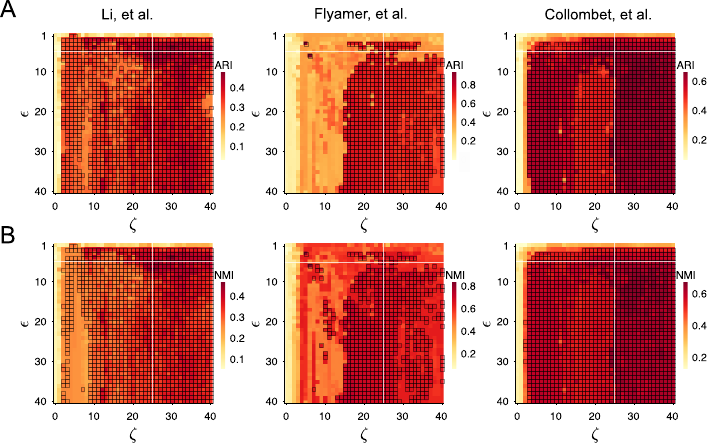}
\caption{The clustering accuracy (ARI in (A) and NMI in (B)) for the SCUDDO algorithm for each of the three datasets, (left) Li, et al.~\cite{bib101}, (middle) Flyamer, et al.~\cite{bib103}, and (right) Collombet, et al.~\cite{bib118}, plotted as a function of the number of sampled superdiagonals $\zeta$ and the embedding dimension $\epsilon$. The pixels in the heatmap are outlined when the ARI or NMI for SCUDDO exceed those of the next best performing method, which is scHiCluster in all cases. The faint horizontal and vertical white lines in each heatmap indicate the row and column for the default values for the SCUDDO algorithm, $\zeta=25$ and $\epsilon=5$.}
\label{fig4}   
\end{figure*}

We next show that SCUDDO can accurately embed single-cell Hi-C maps using a reduced amount of information for already highly sparse single-cell Hi-C maps, surpassing the accuracy of previous algorithms using fewer intrachromosomal matrices for each cell, as well as fewer sampled superdiagonals for each matrix. We study the performance of SCUDDO after restricting the single-cell Hi-C data available to it in two ways: first by varying the hyperparameter $\zeta$ for the number of superdiagonals to sample for each intrachromosomal Hi-C map, as well as varying the hyperparameter $\epsilon$ for the embedding dimension of $\mathcal{V}$.  In Fig.~\ref{fig4}, we show heatmaps of the ARI and NMI for the SCUDDO algorithm, while varying $0 \le \zeta \le 40$ and $1 \le \epsilon \le 40$. The pixels in the $\zeta$-$\epsilon$ plane outlined in black indicate ARI or NMI values for the SCUDDO algorithm that exceed those for all other methods (when they use all of the available single-cell Hi-C data). 

For the Li, et al. dataset, we show in the left column of Fig.~\ref{fig4} (A) and (B) that for $\approx 84\%$ and $82\%$ of the $\zeta$-$\epsilon$ plane SCUDDO outperforms all methods in ARI and NMI. In particular, when $\zeta>2$, SCUDDO gives mean ARI and NMI values over all $\epsilon$ (including low-dimensional embeddings) that match the ARI and NMI for the next best method (when they use all available single-cell Hi-C data). Similarly, when $\epsilon > 1$, SCUDDO gives mean ARI and NMI values over all $\zeta$ that exceed the values for all other methods. Even in regimes with low diagonal sampling and embedding dimension, SCUDDO can obtain ARI values that surpass the next best method (e.g. $\zeta=3,\epsilon= 2$). The maximum ARI and NMI for clustering the  Li, et al. dataset using SCUDDO in the sampled hyperparameter space were  $\approx 0.48$ and $\approx 0.47$ respectively.

On the Flyamer, et al. dataset, SCUDDO outperforms the other methods over a more restricted region of the hyperparameters $\zeta$ and $\epsilon$, as shown in the middle column of Fig.~\ref{fig4} (A) and (B), with $\approx 50\% \text{ and} \approx 43\%$ of $(\zeta,\epsilon)$ input pairs into SCUDDO resulting in ARI and NMI scores that surpassed the next best method's ARI and NMI scores respectively. We find that unlike the other two datasets, SCUDDO requires generally higher $\zeta$ values (more sampled diagonals) to perform state of the art for the Flyamer, et al. dataset. We find that when $\zeta>16$ (across all $\epsilon$) and when $\epsilon>16$ (across all $\zeta$), SCUDDO achieves a larger mean ARI and NMI than the next best method. SCUDDO also achieves exceptional accuracy at $\epsilon=5$, $\zeta=3$ and $\epsilon=7$, $\zeta=6$ with ${\rm ARI} \approx 0.93$ and $0.94$ and ${\rm NMI} \approx 0.80$ and $0.81$ respectively. 

For the Collombet, et al. dataset in the right column of Fig.~\ref{fig4} (A) and (B), SCUDDO outperforms the next best method in ARI and NMI over $\approx 87\%$ and $\approx 90\%$ of the $\zeta$-$\epsilon$ plane.  For $\zeta>3$, the mean ARI and NMI across all $\epsilon$ values for SCUDDO is larger than the other tested methods. Similarly, SCUDDO outperforms the other methods in mean ARI and NMI when $\epsilon > 3$ (across all $\zeta$). Across the sampled hyperparameters, we find that the maximum ARI and NMI are $\approx 0.66$ and $ \approx 0.67$ respectively. 

Previous single-cell clustering algorithms often require a large number of dimensions ($\epsilon \gtrsim 100$) to achieve reasonable clustering accuracy on single-cell Hi-C maps~\cite{bib125}. In addition, the ARI and NMI can possess large fluctuations as a function of the embedding dimension and depend strongly on the specific dimensionality reduction technique that is implemented, for instance with some methods requiring specific dimensionality reduction techniques to be competitive~\cite{bib124}. In contrast, we have shown that the ARI and NMI scores for the SCUDDO algorithm are large at both very low embedding dimensions and when sampling only a few superdiagonals. This result is true even when we treat $\epsilon$ as the final embedding dimension of the output for SCUDDO, despite the fact that SCUDDO always outputs a $l$ dimensional embedding, where $l\leq \epsilon$ for all datasets studied. In addition, we find that SCUDDO does not depend sensitively on the specific dimensionality reduction technique. For instance on the Collombet, et al. dataset, SCUDDO performs roughly equivalently when $\mathcal{V}$ is embedded spectrally (i.e. the default embedding) (${\rm ARI}\approx {\rm NMI} \approx 0.64$), embedded using UMAP~\cite{bib113} (${\rm ARI}\approx {\rm NMI} \approx 0.60$), embedded using t-SNE~\cite{bib112}  (${\rm ARI}\approx {\rm NMI} \approx 0.57$), and when there is no further dimensionality reduction and using $\mathcal{V}$ directly (${\rm ARI} \approx {\rm NMI} \approx 0.56$). While the default values for $\zeta$ and $\epsilon$ for SCUDDO were not optimized to the three selected datasets, the default parameters give excellent results for ARI and NMI for these datasets. However, there are $(\zeta, \epsilon)$ pairs, e.g. $\zeta=26, \epsilon=7$, that give superior performance across all datasets in this manuscript.

\begin{figure}[!t]%
\centering
\captionsetup{width=.9\linewidth}
\includegraphics[scale=1]{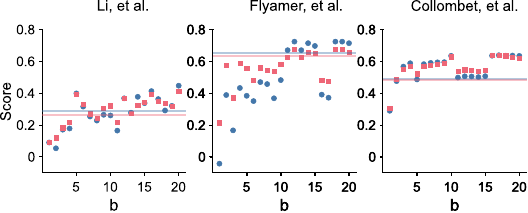}
\caption{The clustering accuracy, ARI (red squares) and NMI (blue circles), for the SCUDDO algorithm plotted as a function of the number of sampled chromosomes $b$ for the three datasets: (left) Li, et al.~\cite{bib101}, (middle) Flyamer, et al.~\cite{bib103}, and (right) Collombet, et al.~\cite{bib118}.  The faint horizontal red and blue lines represent the values of ARI and NMI for the scHiCluster method, which is the next best performing method for these datasets.}
\label{fig5}   
\end{figure}

In Fig.~\ref{fig5}, we calculate ARI and NMI for the three datasets versus the number of intrachromosomal maps $b$ that we sample. For these calculations, we sample all chromosomes with an index less than or equal to $b$, for instance if we set $b=4$, the SCUDDO algorithm samples only chromosomes $1$, $2$, $3$, and $4$. For the Li, et al. and Collombet, et al. datasets in the left and right panels, we find that the ARI and NMI for the SCUDDO algorithm first exceed those for the next best method when $b \gtrsim 4$ and $2$, respectively. However, for the Flyamer, et al. dataset in the center panel, 
most of the chromosomes are needed to achieve high accuracy, with comparable performance with the next best method at $b =11$. We also note that the ARI and NMI for the Flyamer, et al. dataset fluctuates more than the values for the other datasets. For instance,  there are large step changes in the ARI and NMI in Fig.~\ref{fig5} (B) for $b=16$ and $17$.

\section{Discussion}

In this article, we develop a novel algorithm, SCUDDO, to determine a low-dimensional representation and then cluster single-cell Hi-C maps. We focused on three difficult-to-cluster single-cell Hi-C map datasets, where the datasets include ground-truth labels for each single-cell Hi-C map. We compared the ARI and NMI metrics for clustering accuracy from the SCUDDO algorithm to those from three other clustering algorithms that were the most accurate in previous single-cell Hi-C map clustering benchmarking studies \cite{bib106,bib128}. The SCUDDO algorithm is for all cases more accurate in terms of both ARI and NMI compared to the other three methods for all datasets. We also find that the SCUDDO algorithm can accurately cluster single-cell Hi-C maps using a fraction of the intrachromosomal Hi-C maps and their diagonals, as well as fewer embedding dimensions.

The SCUDDO algorithm has several advantages over other existing methods for clustering single-cell Hi-C maps. First, SCUDDO uses, to our knowledge, a new and relatively simple imputation technique for single-cell Hi-C maps, smoothing over local neighborhood features in single-cell Hi-C map and then using backwards diffusion using a 2D Gaussian kernel followed by a diffusion operator. This technique improves the accuracy for datasets where there are few cells (e.g. the Li, et al. and Flyamer, et al. datasets), since it both short-range and long-range information transfer within an intrachromosomal Hi-C map. Additionally, SCUDDO mainly uses PCA and MDS for dimensionality reduction, both of which are much more interpretable than dimensionality reduction techniques like UMAP and t-SNE or using complex inscrutable networks that require training to find features. 

The SCUDDO algorithm combines two key features: the diffused (normalized) diagonals of each single-cell Hi-C map and the trinarized differences along the diffused diagonals. The algorithm then calculates the angles between these features for each diagonal (using cosine similarity) and performs several steps of dimensionality reduction to achieve a final embedding. We find that for some datasets both features are necessary to achieve the best clustering accuracy, e.g. for the Flyamer et al, dataset using only one feature scores at best an ARI of only $0.5$. While the details of the features and SCUDDO algorithm are easy to interpret mathematically, the biophysical interpretation of these features is less clear. For example, it is unclear whether the diffusion and smoothing steps used by SCUDDO have a clear biophysical interpretation.

Interesting future studies involve coupling molecular dynamics simulations of polymer models of chromosomes~\cite{bib131} with the SCUDDO algorithm to further improve clustering accuracy and to better understand the biophysical mechanisms that support the efficacy of the methods used in the SCUDDO algorithm. In addition, the SCUDDO algorithm can be applied to synthetic datasets with labels with tunable noise and sparsity, as well as to experimental datasets without labels to predict cell fate.

\section{Competing interests}
No competing interest is declared.

\section{Author contributions statement}

L.M developed and implemented the SCUDDO algorithm to embed and cluster single-cell Hi-C maps and wrote the first draft of the manuscript. M.D.S and C.S.O edited the manuscript and provided input on the methodology.

\section{Funding}
This work was supported by the National Science Foundation Grant Nos. 1830904 (L.M. and C.S.O.) and 2124558 (L.M. and C.S.O.).

\section{Acknowledgments}
We thank Parisa A. Vaziri for insightful comments.

\end{document}